\documentclass[runningheads,10pt]{llncs}
\pdfoutput=1

\usepackage{graphicx}
\usepackage{booktabs}
\usepackage{url}
\usepackage{mathtools}
\usepackage{subfig}
\usepackage{algorithmic}
\usepackage{algorithm}

\newcommand{\Florida}{State 1}
\newcommand{\Michigan}{State 2}
\newcommand{\Colorado}{State 3}

\begin{document}

\title{AnonTokens: tracing re-identification attacks through decoy records}

\author{%
  {Spiros Antonatos} \and
  {Stefano Braghin} \and
  {Naoise Holohan} \and
  {P\'ol MacAonghusa}
}

\institute{IBM Research -- Ireland\\Dublin, Ireland}

\maketitle

\begin{abstract}
  Privacy is of the utmost concern when it comes to releasing data to third parties. Data owners rely on anonymization approaches
to safeguard the released datasets against re-identification attacks. However, even with strict anonymization in place, re-identification attacks
are still a possibility and in many cases a reality. Prior art has focused on providing better anonymization algorithms with minimal loss of information
and how to prevent data disclosure attacks. Our approach tries to tackle the issue of tracing re-identification attacks based on the concept
of honeytokens; decoy or ``bait'' records with the goal to lure malicious users. While the concept of honeytokens has been widely used in the security
domain, this is the first approach to apply the concept on the data privacy domain. Records with high re-identification risk, called \textit{AnonTokens}, are inserted into anonymized datasets.
This work demonstrates the feasibility, detectability and usability of \textit{AnonTokens} and provides promising results for data owners who want to apply our approach to real use cases.
We evaluated our concept with real large-scale population datasets. The results show that the introduction of decoy tokens is feasible without significant impact
on the released dataset.

\end{abstract}

\section{Introduction}

Data owners are concerned about the privacy of their datasets before releasing them to third parties.
The most well-known approach to achieve anonymity in a released dataset is by applying mechanisms that guarantee privacy.
One of the most well known privacy mechanisms is k-anonymity\cite{sweeneykanon}.
Based on the k-anonymity approach, data are generalized and clustered in such a degree that an individual is indistinguishable from at least k-1 other individuals.

Even with anonymization mechanisms in place, there is always a chance to break the anonymity of a dataset.
Either intentionally or unintentionally, a third party might try to perform a re-identification attack and locate individuals inside the anonymized dataset.
Multiple cases of re-identification attacks have been published in the academia and the
press~\cite{Ochoa01reidentificationof,governordeid,netflixreid}. Such cases discourage the owners from sharing future data or forcing them to take legal action.
Although numerous algorithms that guarantee k-anonymity have been proposed \cite{ola,mondrian,GhinitaHilbert}, none of these approaches are able to trace
if a re-identification attack takes place; these approaches focus on providing a low risk chance.

In this paper we propose an approach for inserting decoys to anonymized datasets, and more specifically to k-anonymous datasets.
The main idea is similar to the approach followed by honeypots and honeytokens\cite{honeypotgeneric,honeygen}.
Traditionally, honeypots and honeytokens are being used for security purposes.
They try to lure attackers to target either realistic replicas of production environments or mine decoy documents or records that contain bogus information.
These replicas and decoy tokens are heavily monitored to detect the presence and source of the attack.
For example, a honeytoken can be an e-mail address that does not belong to a real user but rather is being monitored by the security administrators.

\textit{AnonTokens} are decoy records that have high chance of being re-identified
and at the same time do not belong to the non-decoy records. Decoy records can be extracted from public datasets,
like census datasets or voters lists,  and are unique for each recipient third party.
Since we know the anonymity model that is being used to protect the original data, we can use the same model to identify
which records from the public datasets impose the greater risk for re-identification.
Our approach is driven by two main questions.
First, we explore how we can lure attackers that want to perform re-identification attacks on anonymized datasets to attack specific records.
Second, we want to be able to monitor the presence and source of the re-identification attack.
Our approach has to deal with unique challenges. Anonymized datasets deal with real data and thus the use of artificially generated data, like the honeytokens approach,
is detectable and does not serve our purpose. Second, anonymized dataset are stripped out of personal information, so we cannot make use of attributes, like e-mails, which
can be faked and be monitored.
Finally, the use of real data as ``baits'' needs careful management of the consent and legal aspects.

%
%
We study our approach along three major dimensions. First, we explore if the introduction of decoys is feasible in a realistic setting and we measure the
scalability of our mechanism.
Second, we analyze under which conditions an attacker can detect the presence of the decoy records.
Finally, we study the usability of the anonymized datasets that contain decoy records. We document how the various decoy strategies impact the dataset utility.

\textit{AnonTokens} threat model addresses mostly scenarios where data are being shared with a limited set of recipients. We acknowledge that the scenario of releasing
datasets for public download is not covered by our approach due to identity tracking and scalability issues. We assume that the number of recipients is in the order of few tens
that cannot be fully trusted (e.g. they can intentionally re-identify or they might get hacked). 
For example, the re-identification attack against the Governor of Massachusetts happened from ``hospital data released to researchers by the 
Massachusetts Group Insurance Commission (GIC) for the purpose of improving healthcare and controlling costs``~\cite{governor}. Although this is a case which reached the media and
was published, there are many more cases were data are shared with a closed group of recipients. 

The rest of the paper is organized as follows. Section~\ref{sec:background} provides background information about k-anonymity and honeytokens. Section~\ref{sec:related}
describes the related work while Section~\ref{sec:decoys} is an introduction to the decoy records for anonymized datasets and their properties. Section~\ref{sec:expsetup}
provides details about the experimental setup, followed by the sections that describe the feasibility (Section~\ref{sec:feasibility}), detectability (Section~\ref{sec:detectability})
and scalability (Section~\ref{sec:scalability}). We conclude and summarize in Section~\ref{sec:conclusions}.

\section{Background}
\label{sec:background}

\paragraph{\bf Anonymization, k-anonymity and re-identification attacks}

Based on the data privacy terminology, attributes in a dataset are classified as direct, quasi or sensitive.
Direct attributes are uniquely identifying and are always removed or masked before data release.
Quasi identifiers are sets of attributes that can uniquely identify one or very few individuals.
For example, if we observe the gender attribute in isolation is not uniquely identifying (roughly 50\% of a dataset
would be either male or female), same for a ZIP code (several thousands of people might live in the same ZIP code).
However, if we look attributes in combinations then we can go down to very few individuals. As an example,
the combination of ZIP code plus gender plus birth date can be uniquely identifying (in case of US this combination
can identify 87\% of the population).
Sensitive attributes contain information needed by the recipients, such as researchers, and are released directly. Examples
of sensitive attributes include a disease attribute in a medical dataset.

Anonymization algorithms, such as OLA\cite{ola} and Mondrian\cite{mondrian},
operate on the quasi-identifiers with the goal to generalize them up to a degree that cease to be uniquely identifying.
Other approaches, such as Relaxed Mondrian~\cite{mondrian} and Xu et al.'s Bottom-up
Greedy Algorithm~\cite{xubottomup}, use local recoding that is more flexible as it allows overlapping among different anonymisation groups. In this work,
we do not evaluate our approach with local recoding algorithms.

By observing the dataset of Table~\ref{tbl:quasi}, we notice that the name attribute is uniquely identifying.
If we look the age, gender and ZIP code attributes in isolation, they are not uniquely identifying since there are at least two records for each value.
However, the combination of age plus gender plus ZIP has a unique appearance (``18 - Male - 13121'').

Based on the k-anonymity approach, quasi-identifiers are generalized and clustered in such a degree that an individual is indistinguishable from at least k-1 other individuals.
Generalization is defined by the user. For example, a possible generalization for the gender attribute is ``Male $\rightarrow$ Person, Female $\rightarrow$ Person, Other $\rightarrow$ Person'', which means
that when we want to generalize a specific gender value we will replace it with the value ``Person''. A k-anonymized dataset will have at least k instances for each combination
of quasi-identifiers.

Records with identical values for all quasi-identifiers form an {\it equivalence class}.
As an example, given the anonymized dataset of Table~\ref{tbl:exampleanonymized}, 
records 1 to 4 form an equivalence class and records 5 to 7 form a second equivalence class.
The data anonymization process generalizes the values of quasi-identifiers and partitions the dataset until the privacy
guarantees are met. As an example, let us consider the example dataset of Table~\ref{tbl:quasi}. A k-anonymized view, k = 2,
of the dataset can be seen at Table~\ref{tbl:exampleanonymized}. Note that other anonymized views exist as well. In this example, the age attribute
has been generalized to intervals in the last three records. One can observe that, in the anonymized dataset, all possible combinations of age, gender and ZIP code appear two or more times.

\begin{table}[tb]
  \centering
  \caption{Example dataset with one direct identifier and two quasi-identifiers}
  \subfloat[Original
  \label{tbl:quasi}]{%
  \scriptsize
  \begin{tabular}{|l|l|l|l|l|}
    \hline
    \textbf{Record ID} & \textbf{Name} & \textbf{Age} & \textbf{Gender} & \textbf{ZIP} \\
    \hline
    \hline
    1 & John & 18 & Male & 13122 \\
    2 & Peter & 18 & Male & 13122 \\
    3 & Mark & 19 & Male & 13122 \\
    4 & Steven & 19 & Male & 13122 \\
    5 & Jack & 18 & Male & 13121 \\
    6 & Paul & 20 & Male & 13121 \\
    7 & Andrew & 20 & Male & 13121 \\
    \hline
  \end{tabular}
}
\quad
\subfloat[Anonymized
\label{tbl:exampleanonymized}]{%
  \scriptsize
  \begin{tabular}{|l|l|l|l|l|}
    \hline
    \textbf{Record ID} & \textbf{Name} & \textbf{Age} & \textbf{Gender} & \textbf{Zip} \\
    \hline
    \hline
    1 & A & 18-20 & Male & 13122 \\
    2 & B & 18-20 & Male & 13122 \\
    3 & C & 18-20 & Male & 13122 \\
    4 & D & 18-20 & Male & 13122 \\
    5 & E & 18-20 & Male & 13121 \\
    6 & F & 18-20 & Male & 13121 \\
    7 & G & 18-20 & Male & 13121 \\
    \hline
  \end{tabular}
}
\end{table}

Re-identification attacks~\cite{henriksenreid,elemamreview} aim to break the anonymity of the released datasets.
Based on the classification in~\cite{elemamreview}, there are three main types of attacks.
The first type of attack aims to re-identify a specific person for whom the attack has prior knowledge that they exist in the de-identified dataset. Our approach is not focused on this type of attack.
The second attack aims to identify an individual by using sources of public information about an individual or individuals that are also present in the de-identified dataset.
The last attack involves re-identifying as many people as possible from the de-identified data. The last two types of attacks rely on linking the anonymized datasets with other public or
private datasets. Our approach is useful for tracing these two types of attacks since the decoy records can be used to lure the attention of the attacker to specific records.
The grand challenge with anonymization is that the utility of the original dataset is decreased since data are being generalized and/or suppressed.
The loss in utility is captured through information loss metrics.
Several metrics have been proposed, such as non-uniform entropy~\cite{nonuniformentropymetric}, precision~\cite{precisionmetric},
discernibility~\cite{discernibilitymetric}, average equivalence class size~\cite{mondrian}, generalized loss metric~\cite{glmmetric} and global certainty penalty~\cite{gcpmetric}.
We will come across the information loss metrics in Section~\ref{sec:scalability} where we measure the usability of our approach.

\paragraph{\bf Honeypots and honeytokens}
A formal definition of a honeypot is a ``trap set to detect, deflect or in some
manner counteract attempts at unauthorized use of information systems''\cite{DBLP:journals/ieeesp/Spitzner03}. 
Practically, honeypots are computer systems set up to lure attackers.
They are non-production systems, which means machines that do not belong to any user
or run publicly available services. Instead, in most cases, they passively wait for attackers to
contact them. By default, all traffic destined to honeypots is malicious or unauthorized
as it should not exist in the first place.

A honeytoken is a honeypot which is not a computer. Instead it can be any type of digital entity.
A honeytoken can be a credit card number, a spreadsheet, a PowerPoint presentation, a database entry, or even a bogus e-mail/login.
Honeytokens come in many shapes or sizes but they all share the same concept: a digital
or information system resource whose value lies in the unauthorized use of that resource.
Just as a honeypot computer has no authorized value, no honeytoken has any authorized use.

\section{Decoys for anonymized datasets}
\label{sec:decoys}

In this Section we provide an in-depth description of decoy records for anonymized datasets. The purpose of the decoy records is to lure attackers to
perform re-identification attacks on the decoy records that intentionally have a higher re-identification risk than the rest of the anonymized dataset.
Decoy records need to maintain two basic properties. First, they need to correspond to real records since re-identification attacks are performed
through linking with real public or private datasets. Second, they need to be associated with high re-identification risk and thus correspond
to small enough sets of the population. This property puts a practical limit on the total number of decoy records available to be used.

\begin{figure*}[bt]
  \subfloat[\label{fig:singlenodecoy}]{\includegraphics[width=0.22\textwidth]{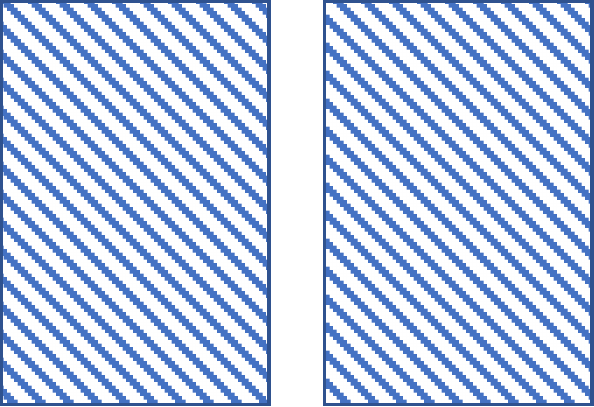}}
  \quad
  \subfloat[\label{fig:multiplenodecoy}]{\includegraphics[width=0.22\textwidth]{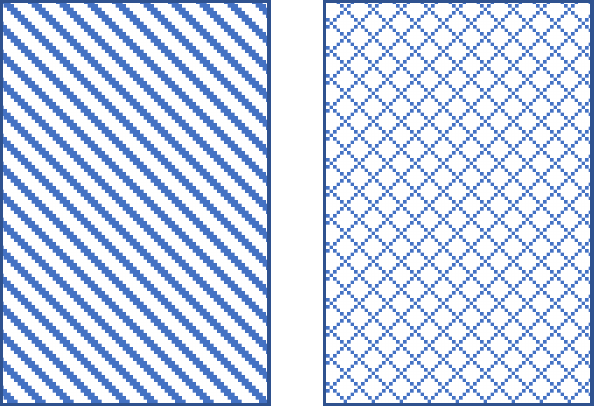}}
  \quad
  \subfloat[\label{fig:multiplewithdecoy}]{\includegraphics[width=0.22\textwidth]{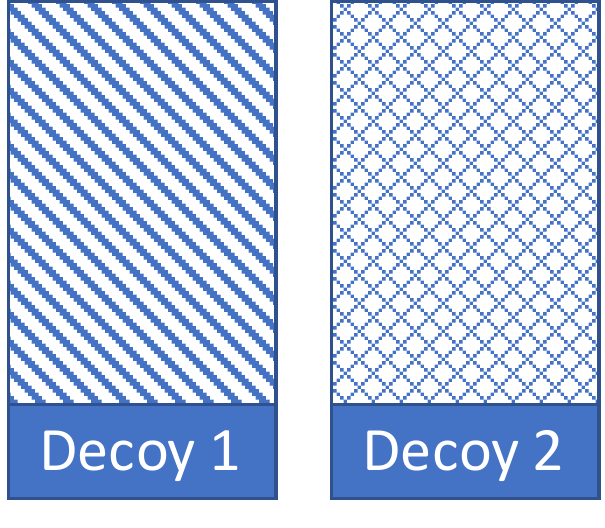}}
  \quad
  \subfloat[\label{fig:anontokens}]{\includegraphics[width=0.22\textwidth]{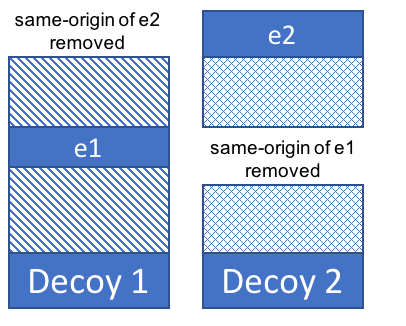}}
  \caption{Possible scenarios of multiple third parties receiving anonymized datasets}
  \label{fig:scenarios}
\end{figure*}

Let's assume that an original dataset D is anonymized to dataset D'. Figure~\ref{fig:scenarios} depicts the possible scenarios when multiple third parties receive anonymized datasets.
In the first scenario (see Figure~\ref{fig:singlenodecoy}) all third parties receive the same D'.
This is the most common scenario in practice.
Tracing back a re-identification attack is not feasible since all parties have the same copy
In this case it is infeasible to trace re-identification attacks back since all parties have the exact same copy.

Figure~\ref{fig:multiplenodecoy} describes the scenario where a different anonymized version of the dataset is shared to each third party.
These versions are based on the original dataset D only and do not contain any decoys.
Previous work has used information loss metrics \cite{kanonfingerprint} to create variations of D' without considering the risk.
Those works introduced fingerprinting to prevent data disclosure attacks but not re-identification attacks.
In an event of a re-identification attack, the fingerprints cannot disclose much about the source of the attack.
Thus, Data disclosure attacks can be detected but not re-identification attacks

In the third scenario, shown in Figure~\ref{fig:multiplewithdecoy}, each third party receives a different variation of D', via the application of fingerprinting and
watermarking methods, and each version is also appended with unique decoy records.

However, fingerprinting\slash watermarking methods rely on applying different generalization levels to the exact same set of original records.
Two or more colluding parties can detect which equivalence classes derive from the same original records and identify the decoy ones.
Re-identification attacks can be detected but decoy records are detectable since they do not have any same-origin class on the other datasets

\begin{definition}
 Two equivalence classes e\textsubscript{1} and e\textsubscript{2} with quasi attributes Q\textsubscript{1}, Q\textsubscript{2}, \ldots, Q\textsubscript{n}
and generalization hierarchies H\textsubscript{1}, H\textsubscript{2}, \ldots, H\textsubscript{n}
originate from the same original records if for every Q\textsubscript{i}, the value of Q\textsubscript{i} in e\textsubscript{1} is either a parent
or a child of the value of Q\textsubscript{i} in e\textsubscript{2} in the corresponding generalization hierarchy H\textsubscript{i}. We call these classes {\it same-origin equivalence classes}. It is trivial to prove
that equivalence classes within the same dataset are not same-origin classes.
\end{definition}

As an example of same-origin equivalence classes, let us assume that the quasi-identifiers are the gender, ZIP code and year of birth attributes
and their generalization hierarchies are the ones described in Table~\ref{tbl:generalizations}. An original record has the values ``Male, 55555, 1981'' and we create
two anonymized datasets with generalization strategies (0, 1, 1) and (0, 2, 0) following the generalization levels described in~\ref{tbl:generalizations}.
The anonymized records will look like ``Male, 5555*, 1980-1982'' and ``Male,  555**, 1981''. We notice that ``5555*'' is a child of ``555**'' and ``1981'' is a child of ``1980-1982'', thus
these two records belong to same-origin equivalence classes.

Let us assume a simple scenario where two parties collude to identify the decoy records.
The first party gets an anonymized dataset D' with a decoy equivalence class $d_1$ and the second party receives an anonymized dataset D'' with a decoy equivalence class $d_2$.
For simplicity reasons, no suppression rate is enforced in the anonymized datasets and thus $|D'| = |D''|$.
Every non-decoy equivalence class of D' will have a same-origin non-decoy equivalence class present in D''. The decoy equivalence class $d_1$ will not have any same-origin
equivalence class from D'' including $d_2$ (since decoy equivalence classes are unique per third party). Thus, by detecting the classes of one party's dataset with no same-origin class on the
second party's dataset, we can isolate the decoy equivalence classes.

Our strategy for protecting against collusion attacks is twofold. First, we can make non-decoy classes have no same-origin classes.
This approach is shown in Figure~\ref{fig:anontokens}.
Given an anonymized dataset D' with a decoy equivalence class $d_1$ and n\textsubscript{p} recipients,
for each recipient select one or more equivalence classes  $e_{n}$ and remove their same-origin equivalence classes from the other $n_p - 1$ anonymized datasets.
Once someone tries to collude with other parties, they will now calculate
two or more equivalence classes that a same-origin equivalence class cannot be found. The removal strategy can be either random, based on the class size
or based on re-identification risk values, for example force other high risk classes to have no same-origin classes on the rest of the datasets.
Non-decoy equivalence classes (e1 and e2) are selected and their same-origin classes are removed from the other datasets.
Re-identification attacks can be detected and decoy records are harder to detect.
This approach is hardened against collusion attacks and we will further explore its properties in Section~\ref{sec:detectability}.
At this point, we should note that there is a possibility that non-decoy classes appear in an anonymized dataset with no same-origin equivalence classes
on the other copies due to suppression (when applicable). Second, we can introduce more than one decoy class per recipient. Since the attacker has no prior knowledge of the original dataset,
it is hard to differentiate if a class is a decoy or it is a non-decoy with its same-origin classes removed.

\section{Experimental setup}
\label{sec:expsetup}

\begin{table*}[tb]
\centering
\caption{Dataset Attributes}
\label{tbl:datasets}
\scriptsize
\begin{tabular}{|l|l|l|l|l|l|}
  \hline
   &\bf \# of rows & \bf Last Updated & \bf Direct & \bf Quasi &  \bf \# of EQ \\
  \hline
  \hline
  State 1 & $~13M$  & 03-2017 & Name, Surname, Address & ZIP, gender, YOB, race &$777,655$  \\
          &   & & E-mail, Phone number & &  \\
  State 2 & $~7M$   & 05-2017 & Name, Surname, Address & ZIP, gender, YOB & $153,092$ \\
  \hline
  State 3 & $~3.5M$ & 06-2017 & Name, Surname, Address& ZIP, gender, YOB & $73,587$\\
   &  &  & Phone number & & \\
  \hline

\end{tabular}
\end{table*}

For our experiments, we assumed that released anonymised datasets contain identifiers that are subset of bigger population datasets.
We used three public datasets containing demographic information as the population datasets to draw samples from.
More specifically, we used the dataset from 
voters lists of three U.S. States, that we do not disclose for confidentiality reasons.
The attributes of the datasets are summarized in Table~\ref{tbl:datasets}. Direct identifiers are considered
the attributes that uniquely identify a person, such as a name or an e-mail. These attributes are typically removed or masked during the
de-identification process. The quasi identifiers refer to the attributes that can be used in combination in order to identify an individual and
will be subject to anonymization. We identified direct and quasi identifiers automatically by applying the techniques described in~\cite{fpvi,ducc} in order to avoid errors on
manual vulnerability classification.

Before proceeding to any experiment, we first removed all directly identifiable information like names and addresses, as listed in Table~\ref{tbl:datasets} (see direct attributes).
For every dataset, we extracted 50 uniformly random samples in order to perform the feasibility, detectability and usability measurements.
We applied the optimal lattice anonymization based on~\cite{ola} for our anonymization purposes. Optimal lattice anonymization provides us the flexibility
to choose different generalization strategies for creating different anonymized versions of the same original data~\cite{kanonfingerprint}. The generalization levels we used during the anonymization process
are described in Table~\ref{tbl:generalizations}. Note that the ZIP codes are the standard 5-digit U.S. postcodes.

\begin{table}[tb]
\centering
\caption{Generalization levels for the attributes anonymized}
\label{tbl:generalizations}
\scriptsize
\begin{tabular}{|l|l|l|}

\hline
	{\bf Attribute} & {\bf \# Levels} & {\bf Description} \\
\hline
\hline
	Gender & 2 & Value, Person \\
	Race & 2 & Value, * \\
	YOB & 5 & Value, 2-yr interval, 4-yr interval, 8-yr interval, * \\
	ZIP code & 6 & Value, XXXX*, XXX**, XX***, X****, ***** \\
\hline

\end{tabular}
\end{table}

\section{Feasibility}
\label{sec:feasibility}

In this Section we explore how feasible it is to introduce decoy records on an anonymised dataset based on observations from real datasets.
The main feasibility criterion is that there are enough records in the population dataset, used by an attacker to link
with the anonymized dataset, that have risk higher than the highest-risk records in the dataset. In order to measure how many risky records
are available to the population dataset, the following steps were followed. Given an anonymized dataset D' and a population dataset P

\noindent {\bf Step 1:} For each equivalence class of D', calculate the number of records linked to P. In our setting, the quasi-identifiers are also
considered as the linking attributes.

\noindent {\bf Step 2:} Calculate the minimum number of linked records, {\it minLink}. The maximum re-identification risk of D' is defined as $r = 1/minLink$

\noindent {\bf Step 3:} Find the equivalence classes of the population dataset P that have size less than {\it minLink} and greater or equal than k.
Algorithm 1 describes the process for discovering these equivalence classes. For every equivalence class E' of the anonymised dataset, we remove
the linked records from P. At the end of the process, we recompute the equivalence classes of what is left from P and then perform the checks
to find which classes have size less than {\it minLink} and greater or equal than k. This approach works uniformly for algorithms that apply
the same generalization level to the entire dataset, like OLA, as well as algorithms that apply different generalization levels per equivalence class, like Mondrian.

\begin{algorithm}[tb]
\caption{\bf \textit{Discovery of EQ classes in the population dataset}}
\label{alg:direct}
\begin{algorithmic}[1]
\scriptsize
\REQUIRE Population dataset P, anonymised dataset D', k, minLink
	\FOR{each equivalence class E' of D'}
	\STATE Let L be the linked records of P to E'
	\STATE P = P - L
	\ENDFOR
	\STATE Let candidates = $\emptyset$
	\FOR{each equivalence class S of P}
	\IF{$|S| < minLink$ and $|S| >= k$}
	\STATE candidates = $candidates \cup S$
	\ENDIF
	\ENDFOR
\RETURN candidates
\end{algorithmic}
\end{algorithm}

The equivalence classes discovered in this step will have higher re-identification risk than the equivalence classes of the anonymized dataset D'
since they link to fewer records than the riskiest equivalence class of D' (with size equal to {\it minLink}).
Decoy records can be selected from these classes. One can, for example, select k or more decoys from classes that have size equal to $minLink/2$ and thus introduce
classes with double re-identification risk of the unmodified anonymized dataset. We answer the question of detectability of high risk classes in Section~\ref{sec:detectability}.

We anonymized each random sample with different values of {\it k} and suppression rate using the optimal lattice algorithm~\cite{ola} and the
generalization hierarchies described in Table \ref{tbl:generalizations}.
We then calculated the number of equivalence classes and records that can be used for selecting decoys following the methodology described above.

Table \ref{tbl:expfeasibility} shows our experimental results. The numbers are averages over the 50 random sample runs and the standard deviation is enclosed
in parenthesis for each measurement. We chose the values 2, 5, 10, 20 and 50 as the {\it k} values and 0, 2, 5, 10 and 15\% as suppression rates.
The {\it minLink} describes the average minimum linked records (Step 2) and the ``{\it $< minLink EQ$}'' column presents the number of equivalence classes
in the population that have size less than {\it minLink} (Step 3). The corresponding number of records for these classes is shown in the column named ``$< minLink Records$''.
With the exception of the cases where we apply 0\% suppression rate,
for the rest of the configurations we notice that there are a few hundred to a few thousand equivalence classes with higher risk than the anonymized dataset that
correspond to thousands of records.
The standard deviation measurements indicate that the behavior is stable across samples (shown in parenthesis next to each measurement).
For example, in the case of the \Florida{} dataset, for {\it k} equal to 5 and suppression rate equal to 10\% we have 2\,779 available equivalence classes
that correspond to 95\,808 records as a pool to select decoy records from.
We also observe that the results are consistent across datasets.
For the case of 0\% suppression rate, the algorithm anonymized the sample close to the maximum generalization levels, making it difficult or impossible to find
classes with double risk as the initial risk is either close to zero or already extremely low. We also notice a drop in scale for high {\it k} values (e.g. {\it k} equal to 50) consistently in all datasets,
and this is also due to the higher levels of generalization applied.

\begin{table*}[tb]
\centering
	\caption{Number of equivalence classes and records in population datasets that are less than {\it minLink}. The standard deviation 
        is in parenthesis}
\label{tbl:expfeasibility}
\resizebox{\textwidth}{!}{
\begin{tabular}{|l|l|l|l|l|l|l|l|l|l|l|}
	\hline
	\multicolumn{2}{|c|}{}  & \multicolumn{3}{c|}{\Florida{}} & \multicolumn{3}{c|}{\Michigan{}} & \multicolumn{3}{c|}{\Colorado{}} \\
	\hline
	k & suppr & minLink & $< minLink EQ$ & $< minLink Records$ & minLink & $< minLink EQ$ & $< minLink Records$ & minLink & $< minLink EQ$ & $< minLink Records$  \\
	\hline
2 & 0 & 350 (174) & 3 (2) & 71 (82) & 147364 (1019K) & 2 (5) & 71 (70) & 1810 (2232) & 4 (7) & 77 (92)  \\
2 & 2 & 7 (2) & 1555 (601) & 6161 (4061) & 10 (5) & 740 (240) & 3384 (2014) & 11 (7) & 483 (170) & 2277 (1468)  \\
2 & 5 & 5 (1) & 2218 (964) & 6931 (4070) & 5 (2) & 1569 (816) & 5311 (3781) & 6 (2) & 1654 (577) & 5890 (3174)  \\
2 & 10 & 3 (1) & 3008 (2341) & 7587 (6750) & 5 (2) & 1493 (911) & 4876 (3966) & 4 (1) & 2533 (1097) & 7128 (4231)  \\
2 & 15 & 2 (0) & 4655 (4138) & 10534 (10388) & 4 (1) & 2595 (1429) & 7315 (4934) & 4 (1) & 2533 (1097) & 7128 (4231)  \\
	\hline
5 & 0 & 131110 (120K) & 0 (1) & 17 (66) & 1902K (3191K) & 1 (1) & 70 (69) & 6927 (10869) & 2 (2) & 68 (80)  \\
5 & 2 & 108 (27) & 1268 (466) & 49240 (21482) & 140 (33) & 639 (48) & 23140 (6943) & 168 (40) & 388 (27) & 16290 (4769)  \\
5 & 5 & 94 (26) & 2352 (246) & 71646 (22681) & 140 (33) & 639 (48) & 23140 (6943) & 168 (40) & 388 (27) & 16290 (4769)  \\
5 & 10 & 89 (18) & 2779 (323) & 95808 (28780) & 109 (27) & 1328 (187) & 53172 (18772) & 96 (24) & 2126 (276) & 71915 (20844)  \\
5 & 15 & 72 (20) & 9627 (1259) & 246427 (76325) & 86 (21) & 2998 (459) & 104649 (35946) & 96 (24) & 2126 (276) & 71915 (20844)  \\
	\hline
10 & 0 & 242409 (0) & 0 (0) & 0 (0) & 3659K (3625K) & 0 (1) & 56 (72) & 12014 (14357) & 1 (1) & 67 (74)  \\
10 & 2 & 504 (88) & 324 (20) & 47827 (10006) & 550 (74) & 190 (103) & 35877 (11406) & 531 (108) & 184 (35) & 28099 (7847)  \\
10 & 5 & 398 (70) & 2339 (131) & 272274 (47635) & 445 (63) & 412 (46) & 76345 (18843) & 490 (96) & 567 (34) & 81040 (15631)  \\
10 & 10 & 398 (70) & 2339 (131) & 272274 (47635) & 414 (55) & 798 (76) & 119984 (31521) & 472 (68) & 432 (29) & 63042 (13506)  \\
10 & 15 & 333 (54) & 3236 (285) & 369464 (92616) & 414 (55) & 798 (76) & 119984 (31521) & 472 (68) & 432 (29) & 63042 (13506)  \\
	\hline
20 & 0 & 242409 (0) & 0 (0) & 0 (0) & 5262K (3244K) & 0 (0) & 32 (52) & 15133 (15592) & 0 (0) & 55 (57)  \\
20 & 2 & 1454 (176) & 196 (8) & 74121 (12124) & 1727 (374) & 107 (21) & 29563 (12666) & 1542 (202) & 111 (4) & 38821 (7280)  \\
20 & 5 & 1227 (172) & 1315 (34) & 390844 (40196) & 1234 (144) & 252 (23) & 139806 (28665) & 1456 (196) & 238 (8) & 106560 (11347)  \\
20 & 10 & 1227 (172) & 1315 (34) & 390844 (40196) & 1213 (117) & 574 (105) & 278123 (63124) & 1303 (179) & 650 (20) & 190298 (25627)  \\
20 & 15 & 1108 (129) & 2671 (80) & 671105 (88673) & 1131 (139) & 775 (56) & 379951 (61939) & 1303 (179) & 650 (20) & 190298 (25627)  \\
	\hline
50 & 0 & 242409 (0) & 0 (0) & 0 (0) & 5262K (3244K) & 0 (0) & 32 (52) & 1145K (1640K) & 0 (0) & 27 (38)  \\
50 & 2 & 4364 (447) & 119 (4) & 126755 (17312) & 4501 (266) & 90 (2) & 87474 (11083) & 4772 (589) & 54 (4) & 49771 (9019)  \\
50 & 5 & 4028 (334) & 254 (10) & 320183 (40544) & 4501 (266) & 90 (2) & 87474 (11083) & 4732 (495) & 56 (1) & 51800 (7714)  \\
50 & 10 & 3968 (308) & 722 (10) & 563872 (39294) & 4501 (266) & 90 (2) & 87474 (11083) & 4371 (577) & 163 (121) & 158766 (121948)  \\
50 & 15 & 3814 (297) & 1064 (316) & 939288 (356937) & 3776 (286) & 499 (18) & 734368 (69058) & 4078 (368) & 302 (7) & 303774 (27657)  \\
	\hline
\end{tabular}
}
\end{table*}

\begin{figure*}[tb]
	\subfloat[\Florida{}, k=2]{\label{florida2}\includegraphics[angle=-90,width=0.30\textwidth]{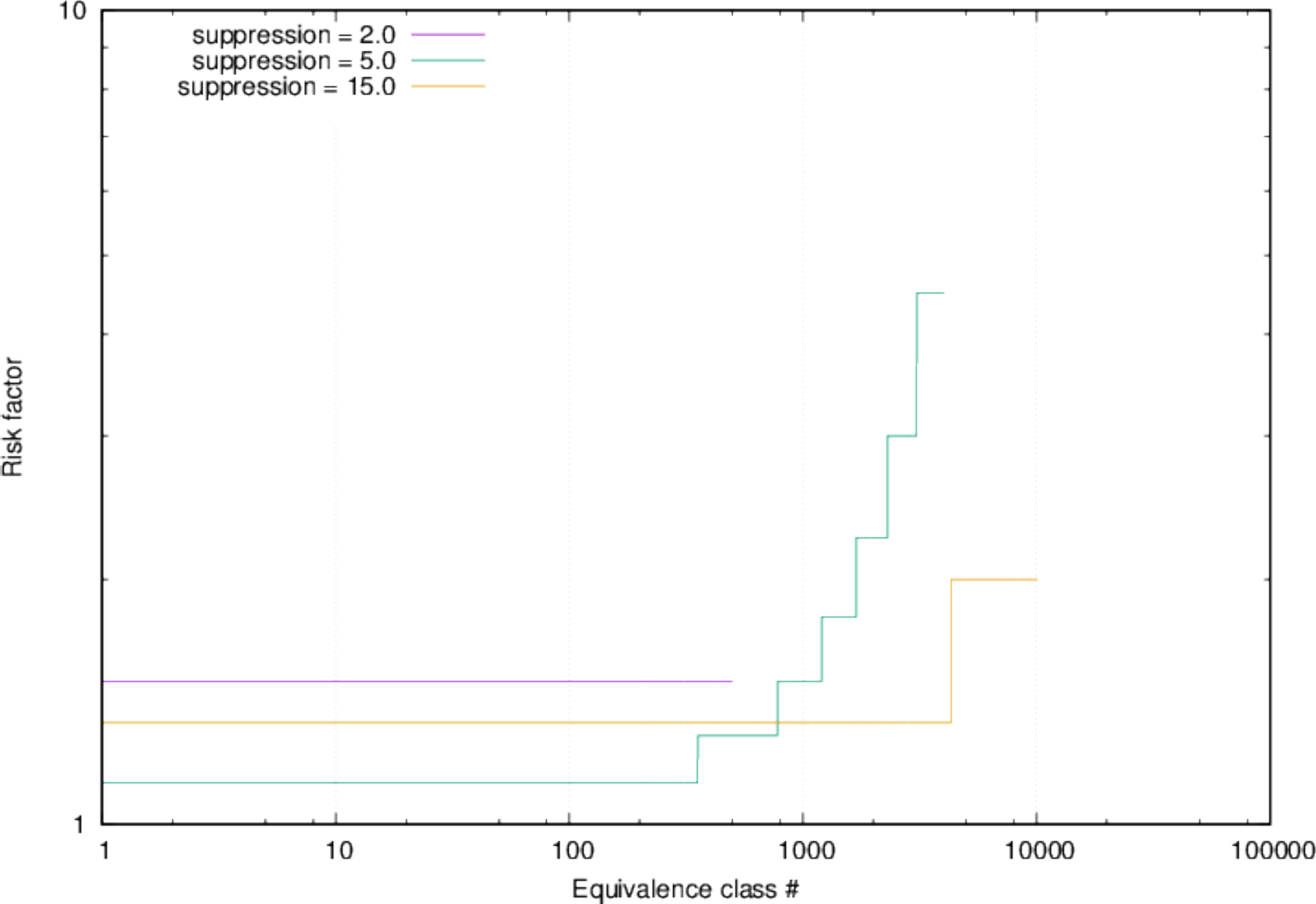}}
  \quad
	\subfloat[\Florida{}, k=10]{\label{florida10}\includegraphics[angle=-90,width=0.30\textwidth]{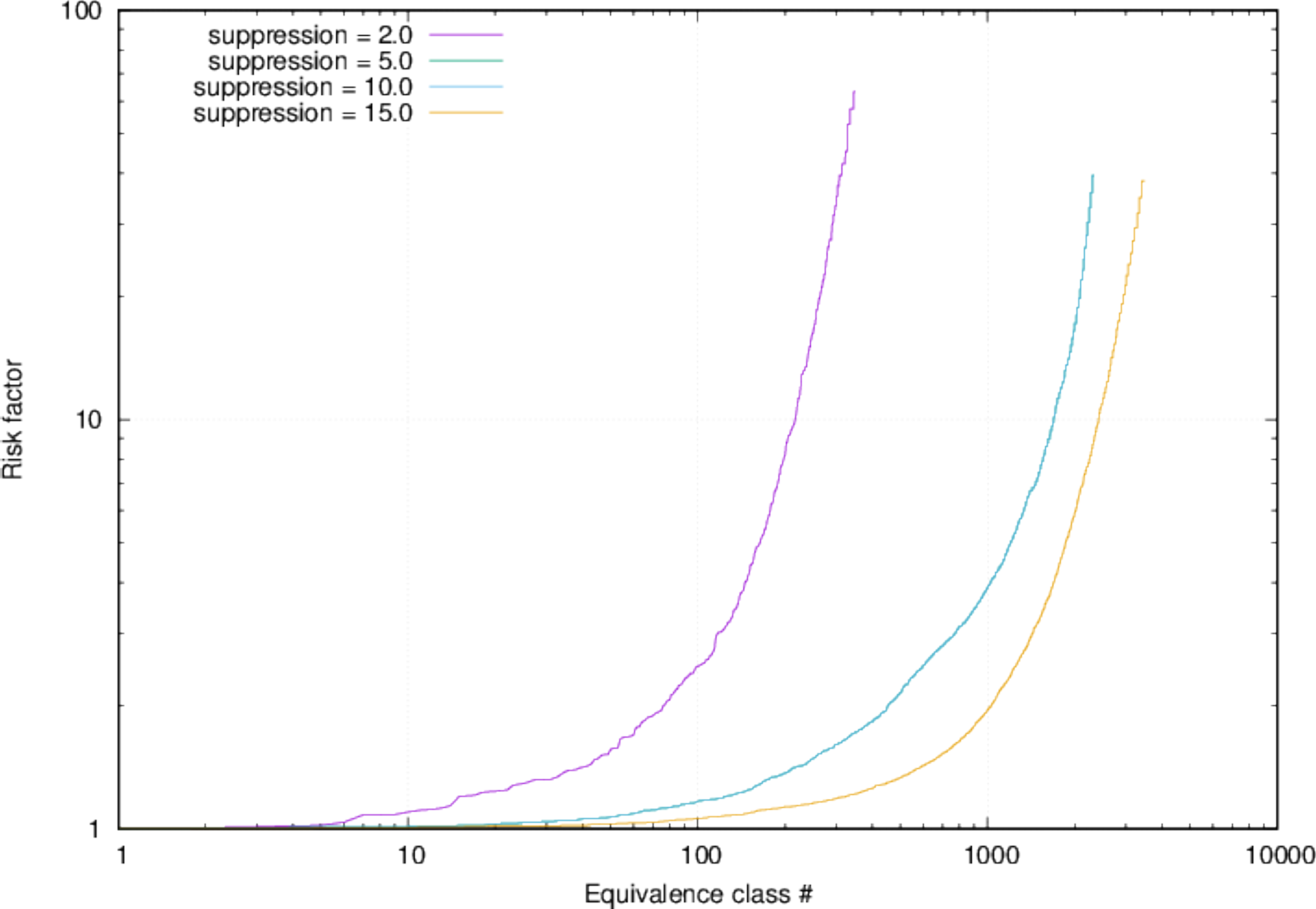}}
  \quad
	\subfloat[\Florida{}, k=20]{\label{florida20}\includegraphics[angle=-90,width=0.30\textwidth]{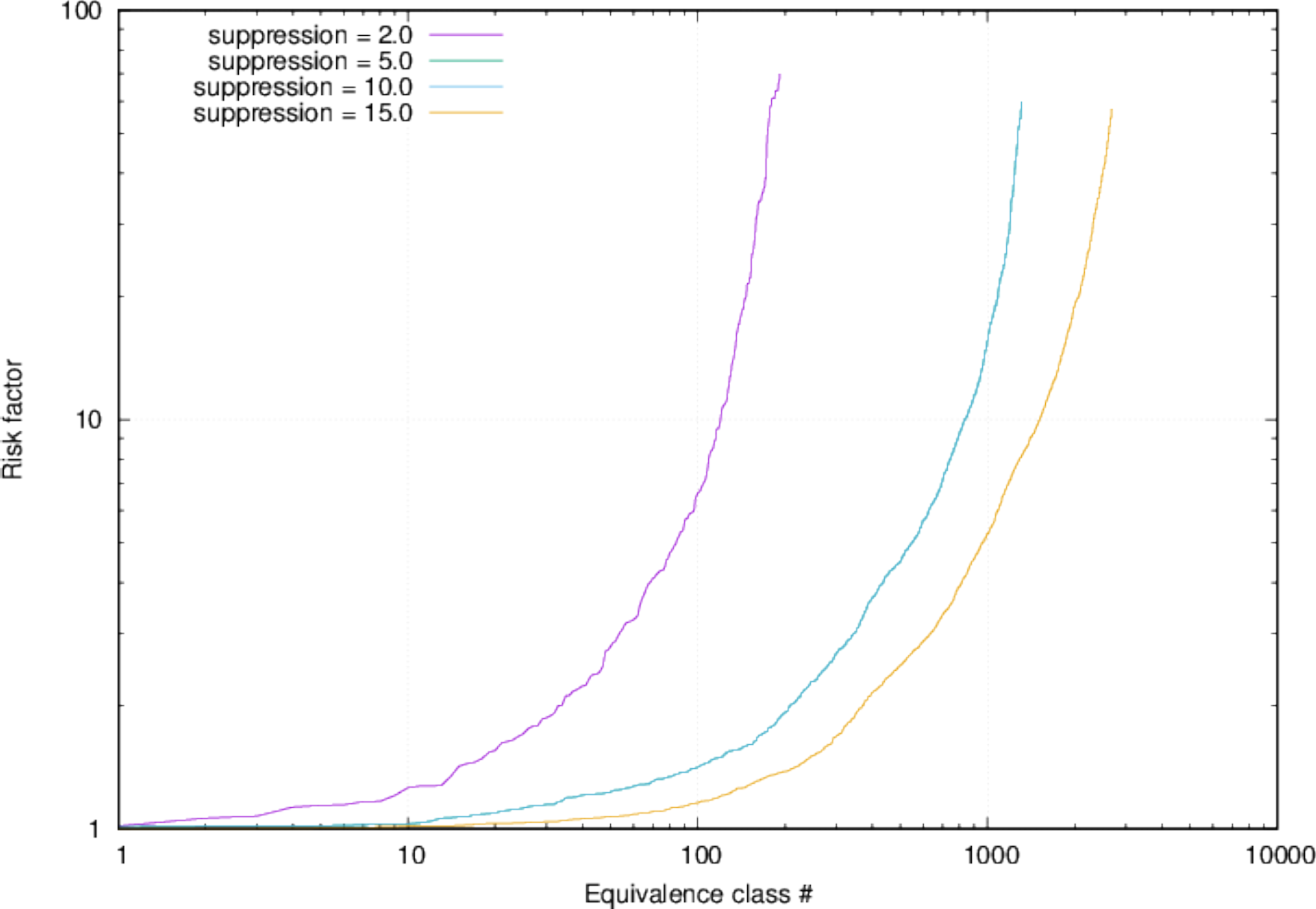}}

	\subfloat[\Michigan{}, k=2]{\label{michigan2}\includegraphics[angle=-90,width=0.30\textwidth]{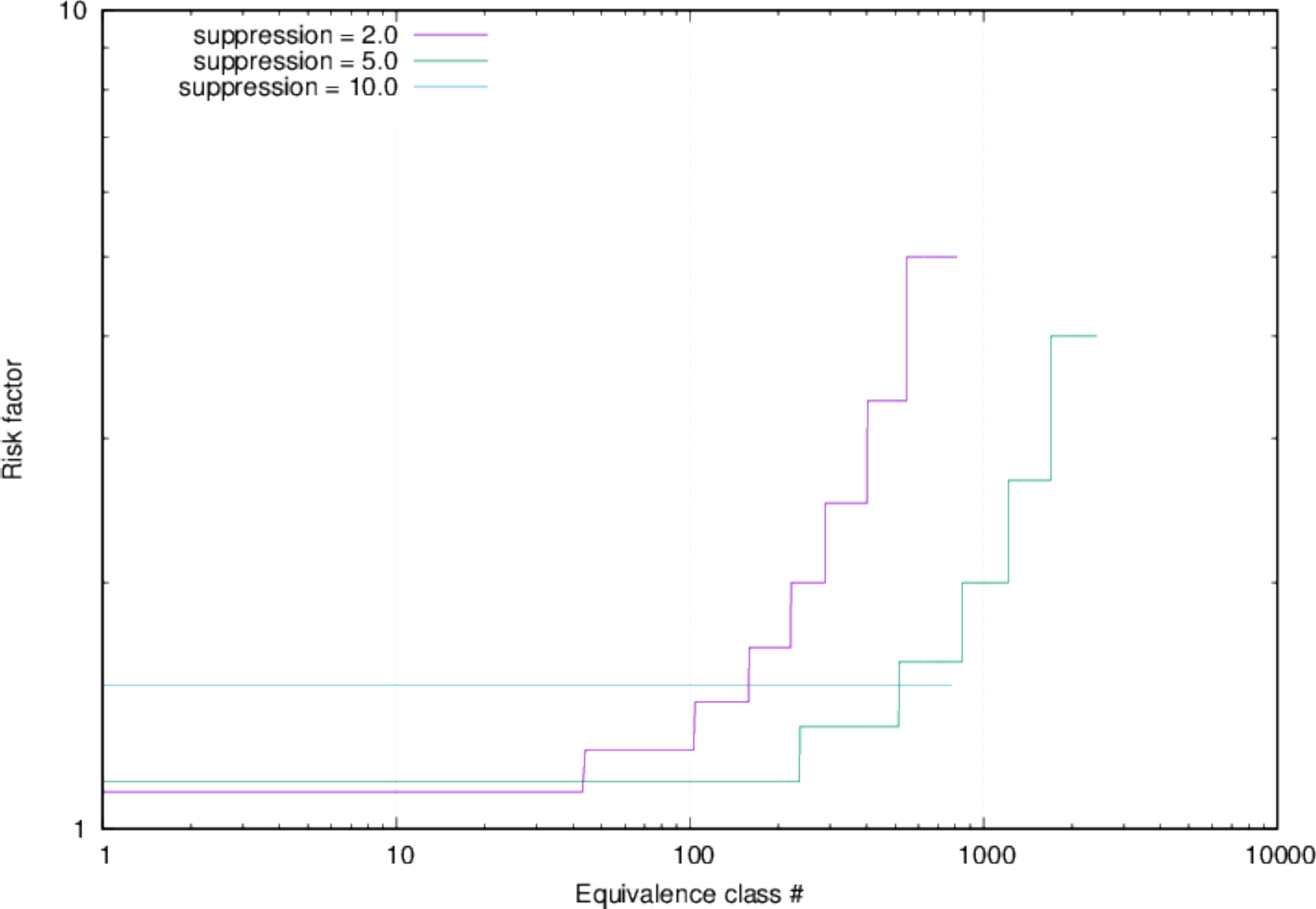}}
  \quad
	\subfloat[\Michigan{}, k=10]{\label{michigan10}\includegraphics[angle=-90,width=0.30\textwidth]{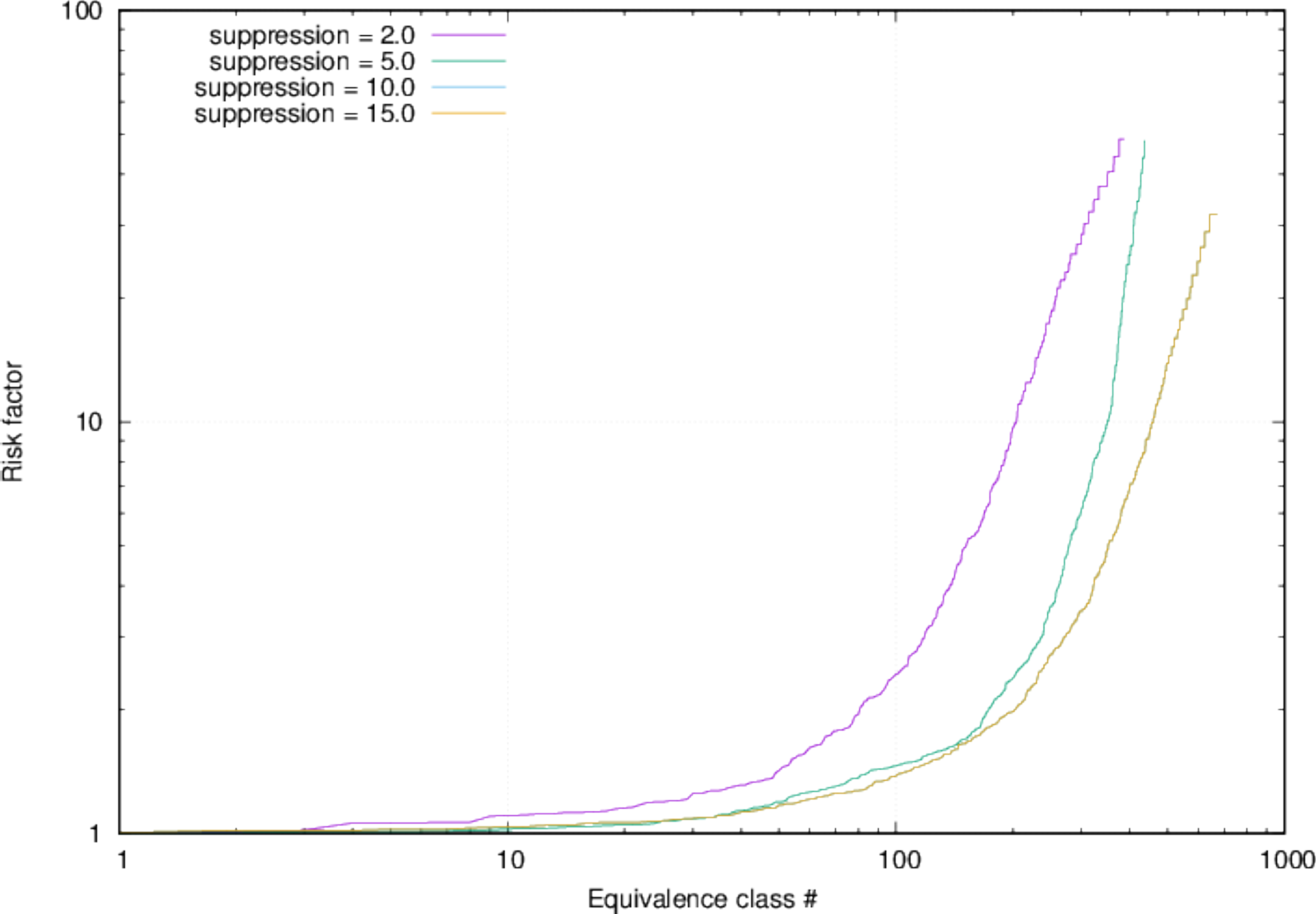}}
  \quad
  \subfloat[\Michigan{}, k=20]{\label{michigan20}\includegraphics[angle=-90,width=0.30\textwidth]{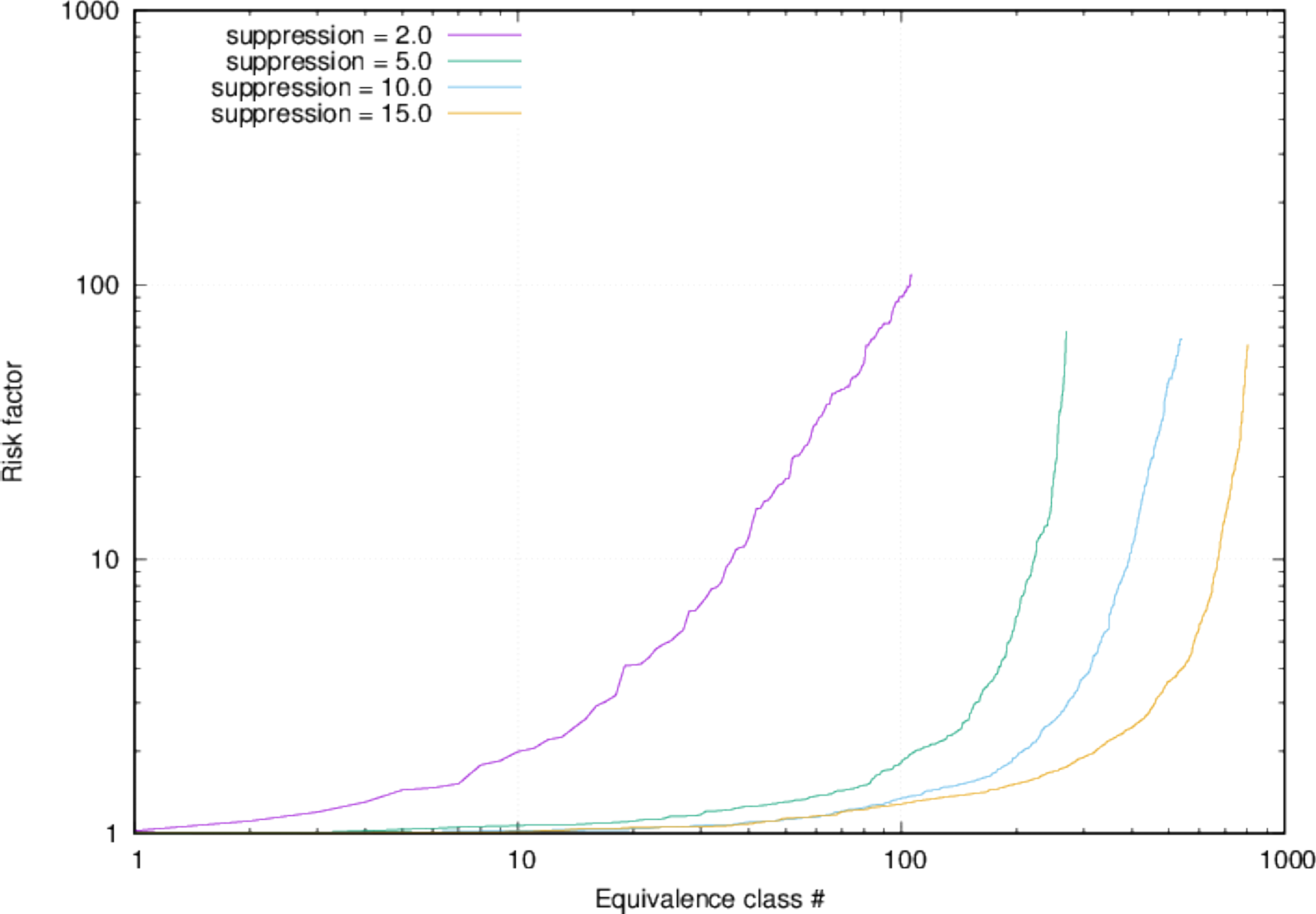}}

 \subfloat[\Colorado{}, k=2]{\label{colorado2}\includegraphics[angle=-90,width=0.30\textwidth]{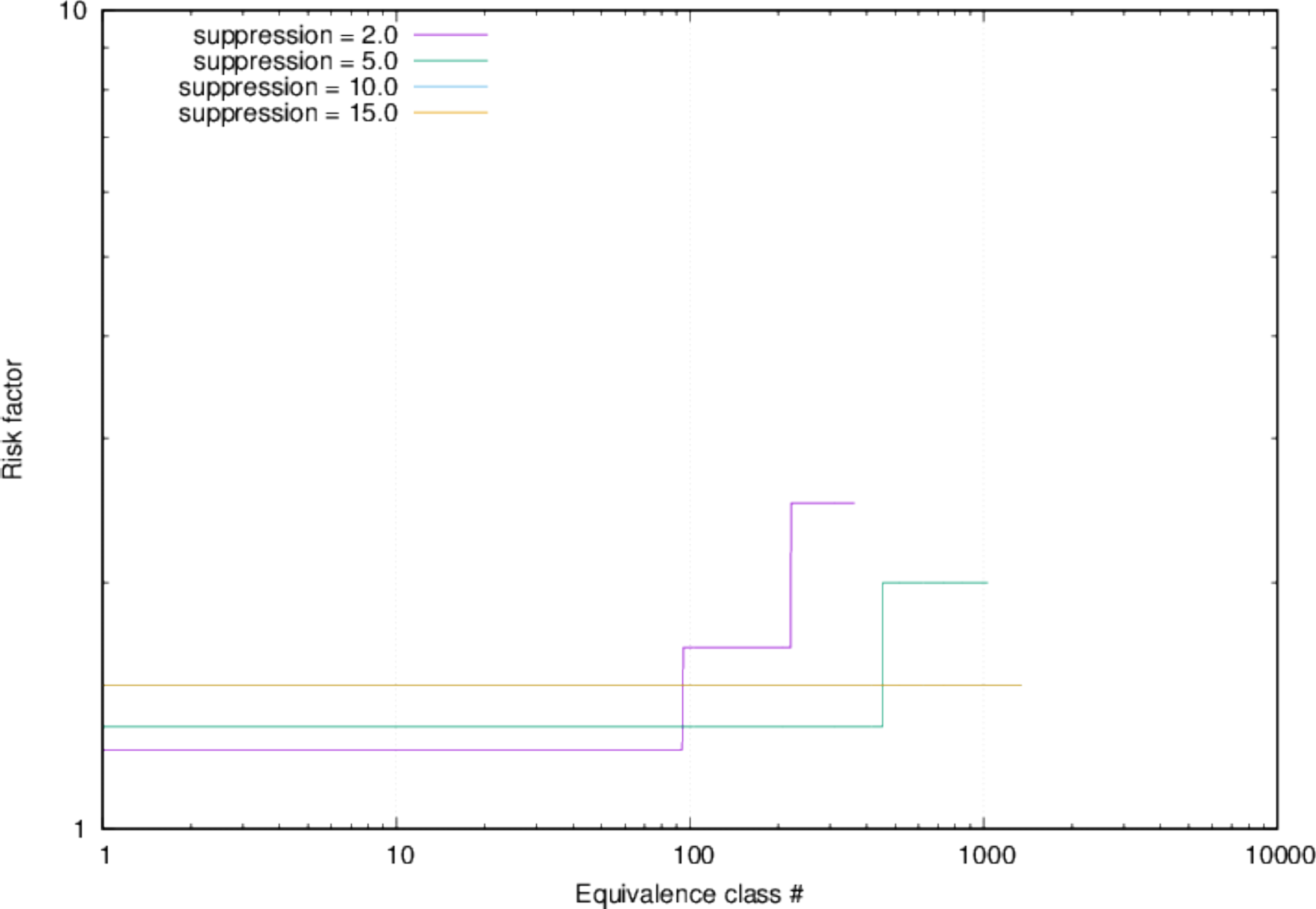}}
  \quad
  \subfloat[\Colorado{}, k=10]{\label{colorado10}\includegraphics[angle=-90,width=0.30\textwidth]{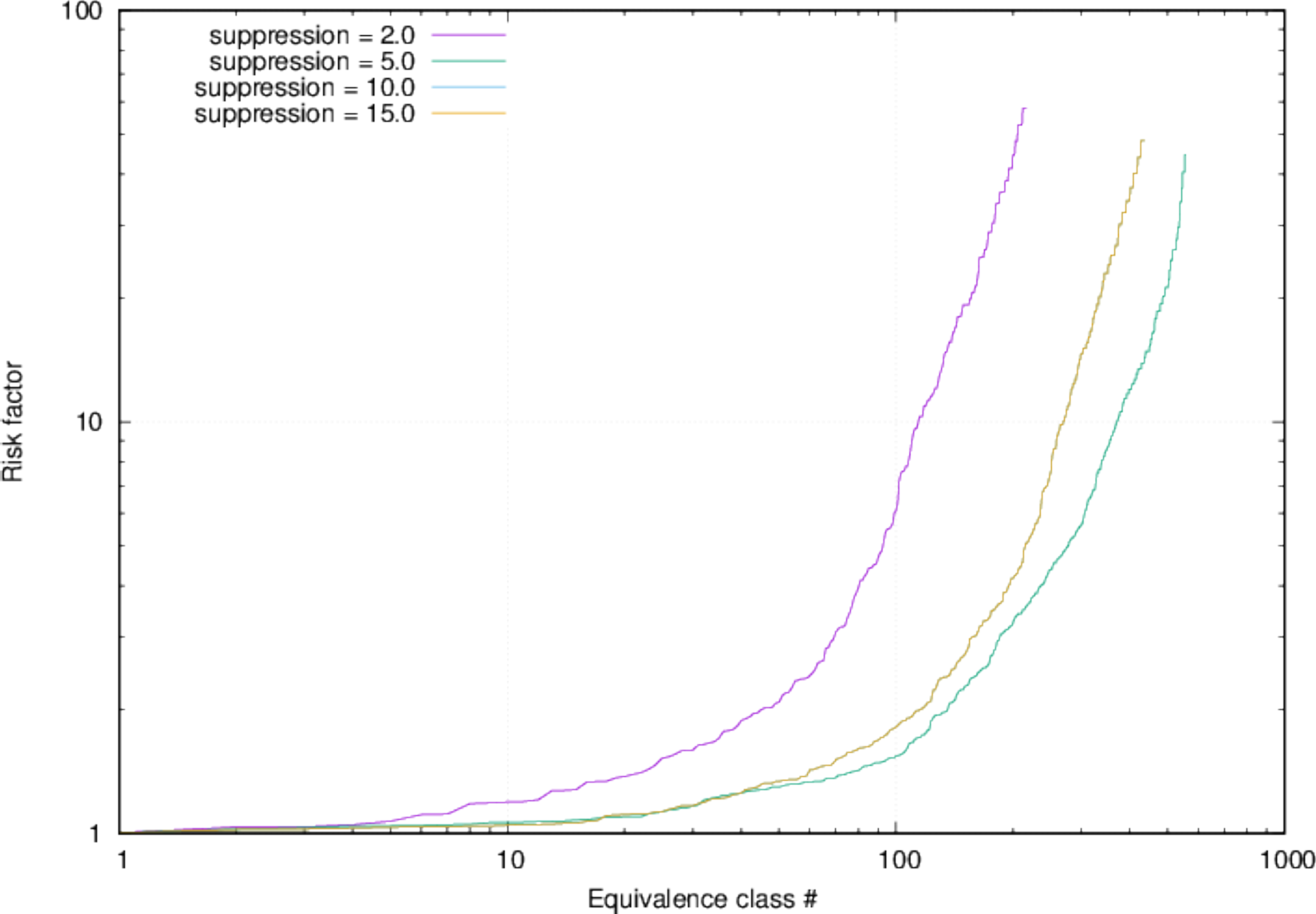}}
  \quad
  \subfloat[\Colorado{}, k=20]{\label{colorado20}\includegraphics[angle=-90,width=0.30\textwidth]{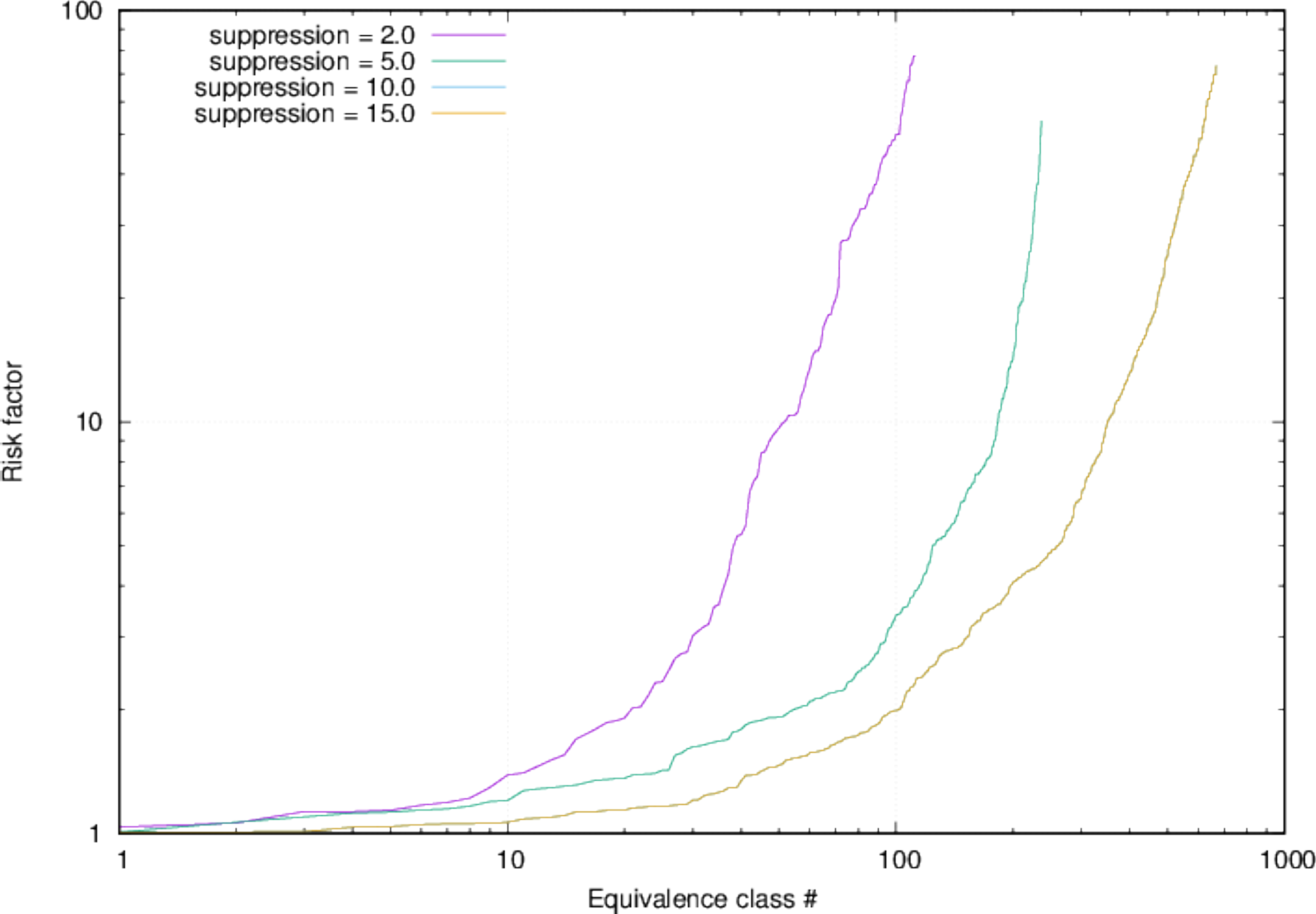}}
	\caption{\label{fig:decoyrisk} Risk multiplication factor for the available datasets and for different values of {\it k} and suppression.
		Both x and y-axis are represented in logarithmic scale.}
\end{figure*}

The question that arises is up to what level of risk we can introduce through the decoy records. For example, can we introduce decoys with double risk chance than
the riskiest equivalence class in the anonymized dataset? We calculated the distribution of the equivalence class sizes that can be used as decoys. Figure~\ref{fig:decoyrisk}
displays the risk multiplication factor for each dataset for various values of {\it k} and suppression. The risk multiplication factor for a decoy class {\it d} is calculated as
$minLink / |d|$. We calculated the risk multiplication factor for every decoy equivalence class and we plotted the risk factors in ascending order. For the {\it k} parameter
we considered the values 2, 5, 10 and 20 while for the suppression rate we selected 2, 5, 10 and 15\%. For space reasons we omit the graphs for k equal to 5 but the results for that
value are very similar to our observation for k equal to 10.

For the \Florida{} case, we observe that for low values of k (e.g k equal to 2) we can introduce decoy records that have up to four times higher risk. For larger values of k, namely 10 and 20,
the risk multiplication factor can be as high as 69.95, meaning that we can introduce records that are close to 70 times riskier than the ones in the anonymized dataset without decoys.
Similar behavior is observed in the cases of \Michigan{} and \Colorado{} datasets, with the maximum multiplication factor being 108.3 and 77.45 respectively. We also observe that for each
setting there are few equivalence classes that have very low risk multiplication factor, with values ranging from 1.1 to 1.2. For example, in the \Florida{} dataset with k equal to 10 and
suppression equal to 10\%, the are, cumulatively, 123 equivalence classes with risk multiplication equal or less than 1.2. This range of multiplication factors allows the user to have better
control of the risk introduced due to the decoy records.

\section{Detectability}
\label{sec:detectability}

In this Section we investigate the detectability aspect of decoy records. There are three basic criteria that need to be fulfilled
in order for the decoy records to be resistant to detection. First, we want to make sure that the sizes of the decoy equivalence classes are not
unique enough to be distinguishable. For example, if the average equivalence class size of an anonymized dataset is 50 and we introduce a decoy equivalence
class of size 10, then it will immediately be noticeable as an outlier. Second, introducing equivalence classes with much higher risk than the rest of the anonymized dataset
might be excluded by an attacker as outliers. If, for example, the riskiest equivalence class in an anonymized dataset without decoys has a 3\% re-identification risk and we introduce
a decoy class with 7\% risk, then this behaviour can be detected and isolated. Finally, decoy records need to be resistant to collusion attacks. In this Section, we
explore the resistance of {\it AnonTokens} to the scenario where multiple attackers collaborate to detect the decoy records.

\begin{figure}[tb]
  \centering
  \includegraphics[angle=-90,width=0.7\columnwidth]{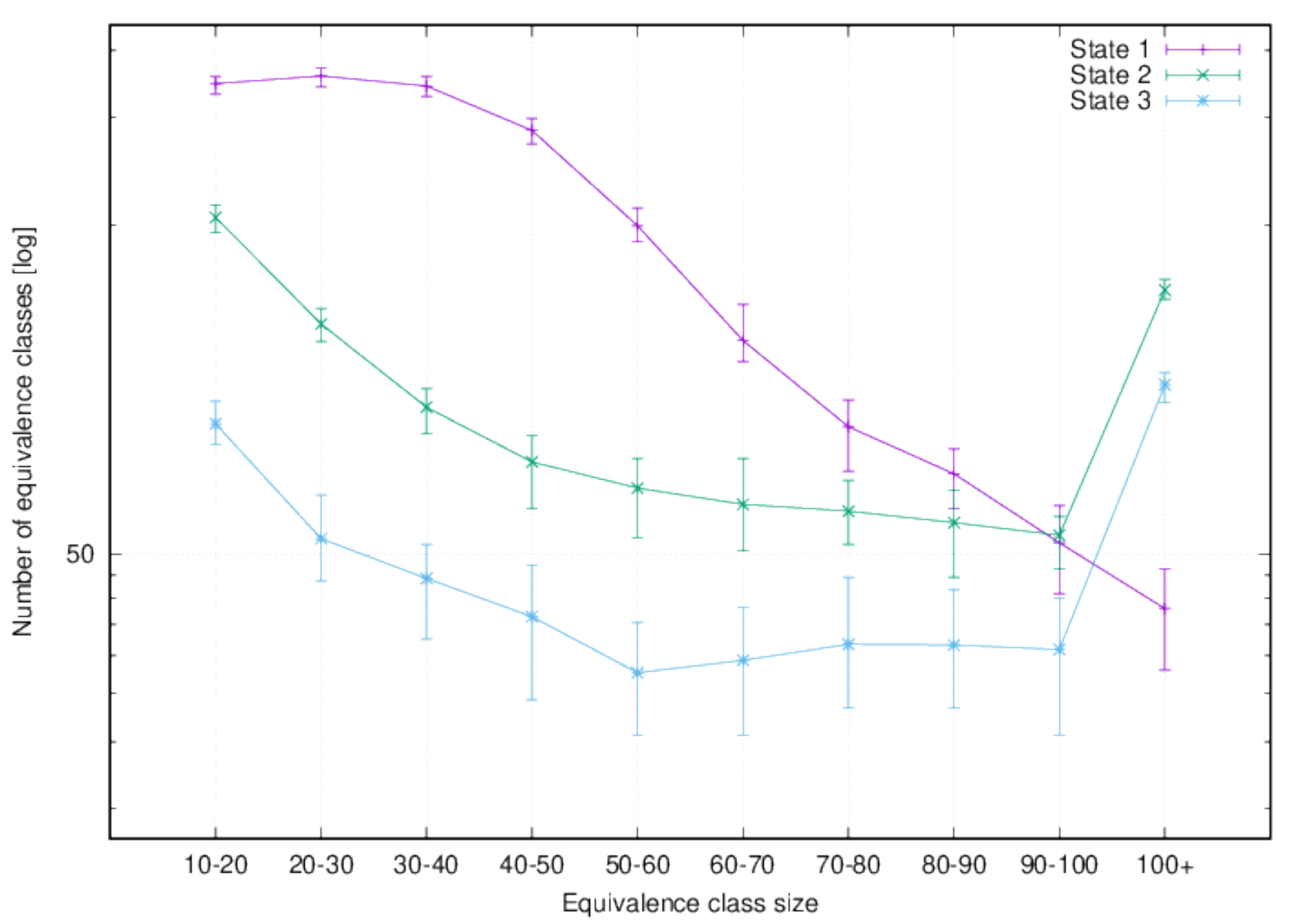}
	\caption{Distribution of equivalence class sizes of the anonymized samples. The {\it k} value was set to 10 and the suppression rate to 5\%. }
  \label{fig:anoneqsizes}
\end{figure}

Regarding the first criterion, we anonymized each random sample with variable {\it k} values and suppression rates and then measured
how many equivalence classes of the anonymized samples have size close to the {\it k} value. We define that an equivalence class Q is close to {\it k} value
if $ k \leq |Q| \leq 1.1\times k$. We select this range of values since attackers will start from the smallest equivalence classes to perform re-identification attacks.
Since anonymized datasets are typically small samples of much larger datasets, large equivalence classes have higher chance to correspond to larger population size,
assuming no bias in the sample extraction.
Table~\ref{tbl:experiments} summarizes the results of our measurements. With the exception of the cases where we applied
0\% suppression rate, we observe that for the rest of the parameters we always find from tens up to thousands of equivalence classes with size close to
the {\it k} value. For example, in the case of the \Florida{} anonymized samples, we can find 135 equivalence class with size close to {\it k} when {\it k} equals 10 and
suppression rate is 5\%. \Michigan{} and \Colorado{} samples have 88 and 26 equivalence classes for these settings. For high values of {\it k}, we notice that few of these equivalence classes exist.
This leads to the conclusion that we can blend in decoy equivalence classes with nearly the smallest size possible as they will be out of place by way of their size.
Even if an attacker ignores the class size and bases their attempts on the re-identification risk, the decoy equivalence classes can blend in. Anonymized samples are
comprised of hundreds or thousands of equivalence classes as Table~\ref{tbl:experiments} indicates (column ``\#EQ in D' '').
Figure~\ref{fig:anoneqsizes} shows the equivalence
class size distribution for {\it k} value equal to 10 and 5\% suppression rate.
We notice that each bin in the distribution includes tens of equivalence classes, with the minimum across
samples being 18 (from the \Colorado{} sample).

As far as the second criterion is concerned, detectability, due to the decoy classes having much higher risk than the rest of the anonymized dataset, is controllable by the data owner.
Decoy classes with significantly higher risk are very attractive to use but at the same time can be isolated as outliers. As shown in Figure~\ref{fig:decoyrisk}, there is a wide range
of risk multiplication factors available for selection. As an example, let us assume that we want to inject a decoy class that is not more than 1.5 times riskier than the rest of the
anonymized dataset. For the \Florida{} case, the minimum percentage of the decoy classes with risk multiplication factor less or equal to 1.5 is 9\% while for the \Michigan{} and \Colorado{} case
these percentages are 5.6\% and 7.6\% respectively. On the other end, if we want to inject decoy classes with more than 4 times higher risk, then 23.7\%, 30.1\% and 47\% of the decoy classes
can be used in the case of \Florida{}, \Michigan{} and \Colorado{} datasets respectively.

\begin{table*}[tb]
\centering
	\caption{Number of equivalence classes in the anonymized samples with size close to the {\it k} value. The standard deviation for each measurement is shown in parenthesis.}
\label{tbl:experiments}
\scriptsize
\begin{tabular}{|l|l|l|l|l|l|l|l|}
	\hline
	\multicolumn{2}{|c|}{}  & \multicolumn{2}{c|}{\Florida{}} & \multicolumn{2}{c|}{\Michigan{}} & \multicolumn{2}{c|}{\Colorado{}} \\
	\hline
	k & suppr. & \#EQ in D' & closeToK  & \#EQ in D' & closeToK & \#EQ in D' & closeToK  \\
	\hline
2 & 0 & 179 (43) & 1 (1) & 90 (69) & 0 (0) & 47 (39) & 0 (1)  \\
2 & 2 & 14236 (40) & 3057 (50) & 4983 (17) & 883 (22) & 2300 (75) & 445 (69)  \\
2 & 5 & 22891 (64) & 8294 (81) & 9436 (231) & 3374 (204) & 3974 (23) & 1210 (30)  \\
2 & 10 & 23161 (62) & 8819 (71) & 12581 (268) & 5536 (240) & 5875 (29) & 2379 (31)  \\
2 & 15 & 31297 (97) & 18517 (81) & 13242 (46) & 6144 (55) & 5875 (29) & 2379 (31)  \\
	\hline
5 & 0 & 105 (86) & 0 (0) & 29 (53) & 0 (0) & 23 (32) & 0 (0)  \\
5 & 2 & 3713 (230) & 211 (29) & 2618 (10) & 231 (9) & 1080 (6) & 110 (7)  \\
5 & 5 & 6614 (22) & 738 (21) & 2618 (10) & 231 (9) & 1080 (6) & 110 (7)  \\
5 & 10 & 9027 (35) & 1845 (35) & 3740 (19) & 633 (22) & 1702 (10) & 263 (13)  \\
5 & 15 & 9917 (38) & 2139 (36) & 4003 (272) & 804 (144) & 1702 (10) & 263 (13)  \\
	\hline
10 & 0 & 25 (0) & 0 (0) & 7 (8) & 0 (0) & 10 (13) & 0 (0)  \\
10 & 2 & 1887 (5) & 31 (5) & 893 (225) & 40 (14) & 537 (47) & 28 (5)  \\
10 & 5 & 3489 (11) & 135 (9) & 1262 (8) & 88 (8) & 511 (4) & 26 (4)  \\
10 & 10 & 3489 (11) & 135 (9) & 2073 (12) & 195 (13) & 818 (9) & 83 (8)  \\
10 & 15 & 4944 (18) & 550 (20) & 2073 (12) & 195 (13) & 818 (9) & 83 (8)  \\
	\hline
20 & 0 & 25 (0) & 0 (0) & 6 (8) & 0 (0) & 8 (6) & 0 (0)  \\
20 & 2 & 945 (3) & 11 (3) & 319 (70) & 3 (4) & 267 (2) & 6 (2)  \\
20 & 5 & 1748 (6) & 48 (7) & 626 (5) & 33 (5) & 254 (2) & 9 (2)  \\
20 & 10 & 1748 (6) & 48 (7) & 1016 (40) & 75 (9) & 406 (4) & 20 (4)  \\
20 & 15 & 2771 (14) & 229 (15) & 925 (7) & 66 (6) & 406 (4) & 20 (4)  \\
	\hline
50 & 0 & 25 (0) & 0 (0) & 6 (8) & 0 (0) & 6 (6) & 0 (0)  \\
50 & 2 & 464 (1) & 5 (1) & 282 (1) & 5 (1) & 133 (15) & 2 (1)  \\
50 & 5 & 837 (3) & 28 (4) & 282 (1) & 5 (1) & 137 (1) & 2 (1)  \\
50 & 10 & 838 (3) & 28 (4) & 282 (1) & 5 (1) & 160 (25) & 5 (3)  \\
50 & 15 & 999 (147) & 65 (34) & 407 (3) & 27 (4) & 186 (2) & 7 (2)  \\
	\hline
\end{tabular}
\end{table*}

For our collusion attack analysis, we assume that $n_p$ recipients receive versions of the anonymized dataset and that all recipients
can collude with each other. We also assume that none of the recipients has knowledge or access to the original dataset, otherwise the introduction of decoys
would be meaningless and their detection would be trivial. For each dataset
we introduce $n_d$ decoy classes and we also select $n_e$ classes to remove their same-origin classes from the rest of the datasets.
In total, we have $n_d + n_e$ classes that are suspected to be decoys. If the attacker knows the values of $n_d$ and $n_e$ then the probability
of a class e being a decoy is $P(e=decoy|n_d,n_e) = n_d / (n_d + n_e)$. However, since attackers has no possible knowledge about these values then
they need to randomly guess if a class is a decoy.

\section{Scalability}
\label{sec:scalability}

In this Section we measure the scalability aspect of {\it AnonTokens} along with their impact on the dataset in terms of information loss.
More specifically, we need to estimate how many recipients we can support with our approach and how many records are removed due to the various protection strategies.
We evaluate two protection strategies for two different cases of modelling the background knowledge of an attacker.

The first protection strategy assumes that an attacker has no prior knowledge of our decoy protection strategy and treats all classes as non-decoys.
In this case, the scalability of our approach is limited by the number of available decoy classes. In the general case, if we have $N_d$ available decoy classes from the population dataset and we choose to
insert a fraction $f$ of them per recipient then we can support up to $1/f$ recipients. In this scenario there is no need to remove non-decoy classes and thus we have
no information loss. On the other hand, since we add information we might alter the use cases of the recipients. If the anonymised datasets are uniformly random samples
of the population, then decoy classes have minimal impact. If the anonymised datasets are biased samples (for example, we want to share datasets for a disease
that affects elder people, such as dementia) then the impact of decoy classes is bigger. In this work, we do not address the issue of bias but it comes down to selecting decoy classes
that are close to the biased sample by calculating the distance of the decoy class to the distance of the sample. For numerical attributes the Euclidean distance can be used,
while for categorical data approaches like that proposed in~\cite{kmodes} can be used. Biased selection of decoy classes affects $N_d$ and thus the fraction $f$.

Our second protection strategy assumes that recipients can collude to discover the decoy classes and thus we need to hide them.
As described in Section~\ref{sec:decoys}, for each participant we select non-decoy equivalence classes and we remove the classes
with the same origin from the datasets of the rest of the recipients. This cascading removal imposes a limit on the number of recipients.
In the general case, if we have $E_d$ available equivalence classes for removal and we need to support $N_r$ recipients, then for each recipient
we can afford to remove up to $E_d / N_r$ non-decoy equivalence classes. However, since the removal process is cascading and we apply it to each recipient,
we might end up removing the entire dataset. For this reason, we introduce a removal budget $b$ and for each recipient we remove up to  $(b \times E_d) / N_r$ non-decoy classes.

We investigate the number of recipients we can support for two removal strategies: a) \emph{random} and, b) \emph{size-based}, where we select classes with size close to the $k$ value.
For our estimation, we assume that the each recipient gets the same anonymized version, thus without any watermarking methods applied where each recipient gets a version
with different generalization levels. This assumption was done for two main reasons. First, watermarking methods impose a strict limit on the number of recipients
since the number of generalization level combinations that can provide information loss close to the optimal one is very low. Second, if a data owner wishes to use a watermarking
mechanism, then the decoy records can be used for fingerprinting. Decoy records are unique per recipient so they can be used to track data disclosure attacks.

The first removal strategy selects classes randomly, independent of their size or risk value. From Table~\ref{tbl:experiments} we can observe that the minimum number
of equivalence classes across all anonymized samples, excluding the case of 0\% suppression rate, is 133 (\Colorado{} case for k equal to 50 and suppression rate set to 2\%).
For other values of $k$ we can see that we have a few hundred to tens of thousands of equivalence classes to select from. Since we have enough equivalence classes to
select from, selecting up to twenty classes for this strategy is feasible. This number can either go up if we need to further reduce the detectability of the decoy classes
or can go down if the decoy classes do not strike out enough in terms of risk. In order to measure the impact on the dataset, we calculated the maximum number of records
removed for ten rounds of the strategy execution for each sample, each value of $k$ and suppression rate. 

The second strategy removes equivalence classes based on their size. From Table~\ref{tbl:experiments} we can observe that there thousands of classes with size equal to the $k$ value.
For example, for State 1, $k$ equal to 5 and a suppression rate of 10\% we can select from 1845 equivalence classes, while the total number of equivalence classes is 9027.
If we want to remove $d$ equivalence classes for a recipient, then this means we also have to remove $d$ same-origin classes from the other recipients. In total, if
K is the number of classes close to the $k$ value, then we can support at most $K/d$ recipients.
Therefore, we can support fewer recipients than the random-based strategy but the number of recipients is prohibitively low. From the previous example, with $K=1845$ and $d=10$ we can support 184 recipients.

A third strategy involving the selection of classes based on their re-identification risk, where the number of high-risk classes is dependent on the sample we draw and the population is possible but we did not consider it for our experiments.

Note that, in practice, the number of recipients is in the order of few tens. Hence, all the presented strategies can easily support this scale of recipients.

\section{Related work}
\label{sec:related}

Sweeney et al.~\cite{sweeneykanon} introduced the concept of k-anonymity and how it can protect against re-identification
attacks via creating indistinguishable records.
El Emam et al.~\cite{ola} proposed a way to achieve globally optimal
k-anonymity. LeFevre et al.~\cite{mondrian} proposed Mondrian as an approach to achieve good balance between generalization and information
loss for multidimensional datasets. These works, along with numerous others that present optimal solution to achieve k-anonymity,
try to prevent re-identification attacks through generalization. They do not introduce any mechanism to trace or redirect re-identification attacks
once they happen. Our approach is orthogonal to all these works; these works can be used to create anonymized datasets with low information loss
and acceptable initial re-identification risk and then our approach appends the decoy records to the output of these algorithms.
Willenborg and Kardaun~\cite{willenborg} identify fingerprints based on combinations of variables in the microdata
records that are short and unique and thus risky. Authors propose anonymity techniques to protect against risky records.
Schrittwieser et al~\cite{kanonfingerprint,genfingerprint} discuss an algorithm to create ``similar'' microdata sets based on data precision. The proposed watermarking
approach applies a k-anonymity clustering algorithm based on data precision and then disseminates different
datasets from the various clusters. The different datasets are generated by using different generalization steps that are close in terms of information loss.
However, such approaches are able to trace data disclosure attacks but not re-identification attacks. All data included in the different versions originate
from the same records so in case a re-identification attack happens the location of source is not feasible.
The approach proposed in this paper is used partly in our approach to generate variations of the anonymized datasets.
Decoy documents~\cite{decoydocuments} propose trap-based defense mechanisms and a deployment platform for addressing the problem of insiders attempting
to exfiltrate sensitive information. The goal is to confuse and confound an adversary requiring more effort to identify real information
from bogus information and provide a means of detecting when an attempt to exploit sensitive information has occurred.
Decoy documents aim at luring attackers that want critical information from systems and does not address the issue of re-identifying individuals in anonymized datasets.
HoneyGen~\cite{honeygen} is a system for generating honeytokens automatically. It creates honeytokens that are similar to the real data by extrapolating the characteristics and properties of real data items.
The honeytoken generation process consists of three main phases: rule mining in which various types of rules that characterize the real data are extracted from the production database;
honeytoken generation in which an artificial relational database is generated based on the extracted rules; and the likelihood rating in which a score is calculated for each honeytoken based on its similarity to the real data.
HoneyGen focuses on generating artificial data and thus it is different from our approach. We differ in two major ways: first, we use real data as honeytokens since we cannot use artificial data. Second, we operate on
anonymized datasets and thus we have to consider the underlying anonymity and risk models to achieve an attractive honeytoken introduction.
Similarly, the work in \cite{baitandsnitch}  discusses situations where decoys are particularly useful as well as characteristics that effective decoy material
should share. Furthermore, authors have developed tools to efficiently craft and distribute decoys in order to form a network of sensors that is capable
of detecting adversarial action that occurs anywhere in an organization’s system of computers.
BotSwindler~\cite{botswindler} is a bait injection system designed to delude and detect crimeware by forcing it to reveal during the exploitation of monitored information.
These works focus only on system security and does not address re-identification attacks on anonymized datasets.

\section{Conclusions and future work}
\label{sec:conclusions}

In this paper we presented a novel approach to identify third parties that try to perform re-identification attacks on anonymized datasets,
and more specifically to k-anonymous datasets, via injection of decoy records. Decoy records is a concept that is widely used in the security domain but this
is the first approach to use the same concept for the data privacy/data sharing domain.

We experimentally demonstrated the feasibility and usability of the approach applying the method to dataset samples drawn from real-world, public datasets. Our results have shown
that the introduction of decoys is doable in the vast majority of the cases and can be applied for tens or even hundreds of recipients. We have also shown the various attacks on
how the decoys can be detected with their respective defences. Even in the case where an attacker suspects the presence of decoys, it is feasible to blend them in with the rest
of the records.

Future work will concentrate on deploying a proof of concept of the approach in a real use case.
This will require a joint work with client's legal department to correctly frame the legal protection for the individuals whose data will be shared in the dataset, and in particular for the individuals represented in the decoy records.
In parallel, disclosure detection and monitoring techniques will require to be further enhanced to not only identify the third party that contributed
to the information disclosure, but also to promptly act upon this type of disclosure.

\bibliographystyle{abbrv}
\bibliography{paper}

\begin{thebibliography}{10}

\bibitem{honeypotgeneric}
Honey pots and honey nets - security through deception.
\newblock
  \url{https://www.sans.org/reading-room/whitepapers/attacking/honey-pots-honey-nets-security-deception-41},
  2001.

\bibitem{netflixreid}
Researchers reverse netflix anonymization.
\newblock \url{http://www.securityfocus.com/news/11497}, 2007.

\bibitem{governor}
The ``re-identification" of governor william weld's medical information'',
  2012.

\bibitem{governordeid}
D.~C. Barth-Jones.
\newblock The "re-identification" of governor william weld's medical
  information: A critical re-examination of health data identification risks
  and privacy protections, then and now, 2012.

\bibitem{discernibilitymetric}
R.~J. Bayardo and R.~Agrawal.
\newblock Data privacy through optimal k-anonymization.
\newblock In {\em ICDE}, 2005.

\bibitem{honeygen}
M.~Bercovitch, M.~Renford, L.~Hasson, A.~Shabtai, L.~Rokach, and Y.~Elovici.
\newblock Honeygen: An automated honeytokens generator.
\newblock In {\em ISI}, pages 131--136. IEEE, 2011.

\bibitem{decoydocuments}
B.~M. Bowen, S.~Hershkop, A.~D. Keromytis, and S.~J. Stolfo.
\newblock {\em Baiting Inside Attackers Using Decoy Documents}, pages 51--70.
\newblock Springer Berlin Heidelberg, Berlin, Heidelberg, 2009.

\bibitem{botswindler}
B.~M. Bowen, P.~Prabhu, V.~P. Kemerlis, S.~Sidiroglou, A.~D. Keromytis, and
  S.~J. Stolfo.
\newblock Botswindler: Tamper resistant injection of believable decoys in
  vm-based hosts for crimeware detection.
\newblock In {\em Proceedings of the 13th International Conference on Recent
  Advances in Intrusion Detection}, RAID'10, pages 118--137, Berlin,
  Heidelberg, 2010. Springer-Verlag.

\bibitem{elemamreview}
K.~El~Emam, E.~Jonker, L.~Arbuckle, and B.~Malin.
\newblock A systematic review of re-identification attacks on health data.
\newblock {\em PLOS ONE}, 6(12):1--12, 12 2011.

\bibitem{ola}
K.~E. Emam, F.~K. Dankar, R.~Issa, E.~Jonker, D.~Amyot, E.~Cogo, J.-P.
  Corriveau, M.~Walker, S.~Chowdhury, R.~Vaillancourt, T.~Roffey, and
  J.~Bottomley.
\newblock A globally optimal k-anonymity method for the de-identification of
  health data.
\newblock {\em JAMIA}, 16(5):670--682, 2009.

\bibitem{GhinitaHilbert}
G.~Ghinita, P.~Karras, P.~Kalnis, and N.~Mamoulis.
\newblock Fast data anonymization with low information loss.
\newblock In {\em Proceedings of the 33rd International Conference on Very
  Large Data Bases}, VLDB '07, pages 758--769. VLDB Endowment, 2007.

\bibitem{gcpmetric}
G.~Ghinita, P.~Karras, P.~Kalnis, and N.~Mamoulis.
\newblock Fast data anonymization with low information loss.
\newblock In {\em VLDB}, 2007.

\bibitem{nonuniformentropymetric}
A.~Gionis and T.~Tassa.
\newblock k-anonymization with minimal loss of information.
\newblock {\em IEEE TKDE}, 21(2), 2009.

\bibitem{fpvi}
A.~Gkoulalas{-}Divanis, S.~Braghin, and S.~Antonatos.
\newblock {FPVI:} {A} scalable method for discovering privacy vulnerabilities
  in microdata.
\newblock In {\em {IEEE} International Smart Cities Conference, {ISC2} 2016,
  Trento, Italy, September 12-15, 2016}, pages 1--8, 2016.

\bibitem{ducc}
A.~Heise, J.-A. Quian{\'e}-Ruiz, Z.~Abedjan, A.~Jentzsch, and F.~Naumann.
\newblock Scalable discovery of unique column combinations.
\newblock {\em Proc. VLDB Endow.}, 7(4):301--312, Dec. 2013.

\bibitem{henriksenreid}
J.~Henriksen-Bulmer and S.~Jeary.
\newblock Re-identification attacks—a systematic literature review.
\newblock {\em International Journal of Information Management}, 36(6):1184 --
  1192, 2016.

\bibitem{kmodes}
Z.~Huang.
\newblock Extensions to the k-means algorithm for clustering large data sets
  with categorical values.
\newblock {\em Data Min. Knowl. Discov.}, 2(3):283--304, Sept. 1998.

\bibitem{glmmetric}
V.~S. Iyengar.
\newblock Transforming data to satisfy privacy constraints.
\newblock In {\em KDD}. ACM, 2002.

\bibitem{mondrian}
K.~LeFevre, D.~J. DeWitt, and R.~Ramakrishnan.
\newblock Mondrian multidimensional k-anonymity.
\newblock In {\em ICDE}, 2006.

\bibitem{Ochoa01reidentificationof}
S.~Ochoa, J.~Rasmussen, C.~Robson, and M.~Salib.
\newblock Reidentification of individuals in {Chicago}'s homicide database: A
  technical and legal study, 2001.

\bibitem{genfingerprint}
S.~Schrittwieser, P.~Kieseberg, I.~Echizen, S.~Wohlgemuth, and N.~Sonehara.
\newblock Using generalization patterns for fingerprinting sets of partially
  anonymized microdata in the course of disasters.
\newblock In {\em 2011 Sixth International Conference on Availability,
  Reliability and Security}, pages 645--649, Aug 2011.

\bibitem{kanonfingerprint}
S.~Schrittwieser, P.~Kieseberg, I.~Echizen, S.~Wohlgemuth, N.~Sonehara, and
  E.~Weippl.
\newblock An algorithm for k-anonymity-based fingerprinting.
\newblock In {\em Proceedings of the 10th International Conference on
  Digital-Forensics and Watermarking}, IWDW'11, pages 439--452, Berlin,
  Heidelberg, 2012. Springer-Verlag.

\bibitem{DBLP:journals/ieeesp/Spitzner03}
L.~Spitzner.
\newblock The honeynet project: Trapping the hackers.
\newblock {\em {IEEE} Security {\&} Privacy}, 1(2):15--23, 2003.

\bibitem{precisionmetric}
L.~Sweeney.
\newblock Achieving k-anonymity privacy protection using generalization and
  suppression.
\newblock {\em Int. J. Uncertain. Fuzziness Knowl.-Based Syst.}, 10(5), Oct.
  2002.

\bibitem{sweeneykanon}
L.~Sweeney.
\newblock K-anonymity: A model for protecting privacy.
\newblock {\em Int. J. Uncertain. Fuzziness Knowl.-Based Syst.}, 10(5), Oct.
  2002.

\bibitem{baitandsnitch}
J.~Voris, J.~Jermyn, A.~D. Keromytis, and S.~J. Stolfo.
\newblock Bait and snitch: Defending computer systems with decoys.
\newblock In {\em Proceedings of the cyber infrastructure protection
  conference, Strategic Studies Institute, September}, 2013.

\bibitem{willenborg}
L.~Willenborg and J.~Kardaun.
\newblock Fingerprints in microdata sets.
\newblock In {\em Statistical Data Confidentiality: Proceedings of the Joint
  Eurostat/UN-ECE Work Session on Statistical Data Confidentiality Held in
  Thessaloniki in March 1999}. Eurostat, 1999.

\bibitem{xubottomup}
J.~Xu, W.~Wang, J.~Pei, X.~Wang, B.~Shi, and A.~W.-C. Fu.
\newblock Utility-based anonymization using local recoding.
\newblock In {\em Proceedings of the 12th ACM SIGKDD International Conference
  on Knowledge Discovery and Data Mining}, KDD '06, pages 785--790, 2006.

\end{thebibliography}

\end{document}